\documentclass[twocolumn,times,tighten]{aastex631}
\usepackage{graphicx,multirow,hyperref,url,color}
\hypersetup{colorlinks=true, urlcolor=cyan, citecolor=cyan}
\newcommand{\mbh}{{M_{\rm BH}}}
\newcommand{\medd}{{\dot{M}_{\rm Edd}}}
\newcommand{\ledd}{{L_{\rm Edd}}}
\newcommand{\msun}{{M_\odot}}
\newcommand{\mdotbh}{\dot{M}_{\rm BH}}
\newcommand{\msunpyr}{{\msun\,{\rm yr}^{-1}}}
\newcommand{\ergs}{{\rm erg\, s^{-1}}}

\begin{document}
\shorttitle{radiative efficiency and direct electron heating}
\title{Observational Constraints on Direct Electron Heating in the Hot Accretion Flows in Sgr A* and M87*}
\shortauthors{Xie et al.}

\author[0000-0001-9969-2091]{Fu-Guo Xie}
\affiliation{Key Laboratory for Research in Galaxies and
Cosmology, Shanghai Astronomical Observatory, Chinese Academy of
Sciences, 80 Nandan Road, \\
Shanghai 200030, People’s Republic of China}

\author[0000-0002-1919-2730]{Ramesh Narayan}
\affiliation{Center for Astrophysics, Harvard \& Smithsonian, 60 Garden Street, Cambridge, MA 02138, USA}
\affil{Black Hole Initiative at Harvard University, 20 Garden Street, Cambridge, MA 02138, USA}

\author[0000-0003-3564-6437]{Feng Yuan}
\affiliation{Key Laboratory for Research in Galaxies and
Cosmology, Shanghai Astronomical Observatory, Chinese Academy of
Sciences, 80 Nandan Road, \\
Shanghai 200030, People’s Republic of China}
\affil{University of Chinese Academy of Sciences, No. 19A Yuquan Road, Beijing 100049, People’s Republic of China}

\begin{abstract}
An important parameter in the theory of hot accretion flows around black holes is $\delta$, which describes the fraction of ``viscously'' dissipated energy in the accretion flow that goes directly into heating electrons. For a given mass accretion rate, the radiative efficiency of a hot accretion flow is determined by $\delta$. Unfortunately, the value of $\delta$ is hard to determine from first principles. The recent Event Horizon Telescope Collaboration (EHTC) results on M87* and Sgr A* provide us with a different way of constraining $\delta$. By combining the mass accretion rates in M87* and Sgr A* estimated by the EHTC with the measured bolometric luminosities of the two sources, we derive good constraints on the radiative efficiencies of the respective accretion flows. In parallel, we use a theoretical model of hot magnetically arrested disks (MAD) to calculate the expected radiative efficiency as a function of $\delta$ (and accretion rate). By comparing the EHTC-derived radiative efficiencies with the theoretical results from MAD models, we find that Sgr A* requires $\delta \ga 0.3$. 
A similar comparison in the case of M87* gives inconclusive results as there is still a large uncertainty in the accretion rate in this source.
\end{abstract}


\section{Introduction}\label{sec:intro}

Accretion flows around black holes (BHs) can be divided into two types: hot accretion flows, which occur at accretion rates below about $1\%$ of Eddington (\citealt*{narayan1994, narayan1995,abramowicz1995}, see \citealt*{yuan2014} for a review)\footnote{Note that the $1\%$ limit quoted here is a typical value for low-luminosity active galactic nuclei \citep{ho2008}. In the case of BH X-ray binaries in the hard state \citep{remillard2006,done2007}, where again hot accretion flows are present \citep{esin1997}, the luminosity can reach much higher values, $\ga10-20\%$ of the Eddington luminosity $\ledd$ (e.g., \citealt{dunn2010}). Here $\ledd = 4\pi Gc m_p \mbh/\sigma_T=1.26\times10^{44}\,(\mbh/10^6\msun)\,\ergs$ for accretion onto a BH with mass $\mbh$.}, and cold accretion disks which are found at accretion rates closer to Eddington \citep{shakura1973, novikov1973,pringle1981}. Most supermassive BHs in the nearby universe have low accretion rates and are believed to have hot accretion flows \citep{ho2008}.

The plasma in a hot accretion flow is two-temperature \citep{shapiro1976}, with electrons and ions having different temperatures. Hence the two species need to be treated with separate energy equations \citep{narayan1995, yuan2014}. The energy equation of electrons takes the form
\begin{equation}
\rho V_r\left({de_{\rm e}\over dR} - {p_e\over\rho^2}{d\rho\over dR}\right) = q_{\rm ie} + \delta q_{\rm vis} - q^-,\label{eq:energy}
\end{equation}
where $\rho$ is the gas density, $V_r$ is the radial velocity, $R$ is the radius, $e_{\rm e}$ and $p_{\rm e}$ are, respectively, the  specific internal energy and pressure of electrons, and $q^-$ is the radiative cooling rate per unit volume.
The first two terms in the right hand side of the above equation correspond to two different mechanisms by which electrons are heated. One mechanism is via Coulomb collisions between ions and electrons, denoted by $q_{\rm ie}$, the other is through direct ``viscous heating,'' which is a catch-all term describing several physical dissipation processes in the plasma as we explain in the next paragraph. We assume that, out of the total viscous heating rate per unit volume $q_{\rm vis}$, a fraction $\delta q_{\rm vis}$ heats the electrons, and the rest $(1-\delta)q_{\rm vis}$ heats the ions. Since almost all the radiation of the accretion flow is emitted by electrons, therefore, when other model parameters such as the mass accretion rate are given, the value of $\delta$ determines the radiative efficiency $\epsilon$ of the accretion flow, which we define as
\begin{equation}
    \epsilon = {L_{\rm bol}\over\mdotbh c^2} = 10\%\,{L_{\rm bol}/\ledd \over \mdotbh/\medd}.\label{eq:eff}
\end{equation} 
Here $L_{\rm bol}$ is the bolometric luminosity of the accretion flow and $\mdotbh$ is the mass accretion rate at the BH horizon. In the second expression, the luminosity and the accretion rate are normalized to, respectively, the Eddington luminosity $\ledd$, and the Eddington accretion rate $\medd = \ledd/(0.1c^2) = 2.21\times10^{-2}\,(\mbh/10^6\msun)\,\msun\,{\rm yr}^{-1}$.

The viscous heating rate of electrons $\delta q_{\rm vis}$ includes several microphysical processes. Many attempts have been made to estimate $\delta$ from first-principle calculations of these processes, including magnetic reconnection \citep{bk1997, quataert1999, ding2010, hoshino2012, hoshino2013, numata2015, sironi2015, rowan2017, rowan2019, ball2018}, magnetohydrodynamic (MHD) turbulence \citep{quataert1998, quataert1999, blackman1999, medvedev2000, lehe2009, howes2010, ressler2015, ryan2017}, dissipation of pressure anisotropy in a collisionless plasma \citep{sharma2007}, or low Mach number shocks \citep{guo2017, guo2018}. Unfortunately, there is no consensus, and the value of $\delta$ remains poorly determined. 

\citet{yuan2003} considered an alternative approach to constrain the value of $\delta$. By calculating the dynamics and radiation of the accretion flow using a height-integrated model, and comparing the results with observations of the supermassive black hole Sagittarius A* (Sgr A*) at the center of our Galaxy, they estimated $\delta \sim 0.5$. 

In this work, we follow the approach of \citet{yuan2003} and estimate the value of $\delta$ using new and updated observational data and a more modern theoretical model. Specifically, we make use of results from the Event Horizon Telescope (EHT), an Earth-size sub-millimeter radio interferometer, which has recently obtained 230 GHz images of the innermost horizon-scale regions of the low-luminosity supermassive BHs, M87* \citep{eht2019, eht2021} and Sgr A* \citep{eht2022a}. Key results from the work of the EHT Collaboration (EHTC) are estimates of the mass accretion rates $\mdotbh$ for the two sources to within a factor of several, see Sec.\ \ref{sec:obs} below. By combining these estimates with a reliable measurement of the bolometric luminosities of the soruces from their broad-band spectra, we are able to evaluate the radiative efficiencies $\epsilon$ (cf. Equation\ \ref{eq:eff}) of the two sources. We compare these measurements with the predictions of the model described in \citet{xie2019} and thereby evaluate the value of $\delta$. 

Before going into details, we briefly introduce additional background information on hot accretion flows. In the last decade or two, it has become clear that the degree of magnetization of the accretion flow plays an important role in the dynamics and observational characteristics of hot accretion flows. Broadly, one distinguishes between two kinds of  models. One has relatively weak magnetic fields and is referred to as the standard and normal evolution (SANE) model, while the other has the maximum saturated level of magnetic field and is called the magnetically arrested disk (MAD) model \citep{narayan2003, igumenshchev2003, bisnovatyi1974, bisnovatyi1976, tchekhovskoy2011, mckinney2012, liska2020}. The dynamical differences between SANE and MAD are investigated in detail in \citet[][see also \citealt{begelman2022,chatterjee2022}]{narayan2012}.

The accumulation of a considerable quantity of poloidal magnetic flux around the BH in the MAD state has the attractive feature that it provides the ideal magnetic field configuration to launch a powerful relativistic jet 
\citep{tchekhovskoy2011, liska2020, narayan2022}. Indeed, there is accumulating evidence that systems with strong relativistic jets are generally MAD, e.g., \citet*{z2014,2014Natur.515..376G,zdziarski2015} for a sample of blazars. Meanwhile, the EHT has provided evidence that M87*  and Sgr A* may also both be MAD \citep{eht2021, eht2022a}. The MAD nature of M87* has been confirmed by an analysis of the observed rotation measure \citep{yuan2022}. These results lead one to speculate that perhaps the MAD configuration is the inevitable final state of most hot accretion flows in nature (e.g., \citealt{narayan2022}).

The radiative properties of MAD flows are found to be fairly similar to those of SANE flows, except that, for a given accretion rate, MAD is brighter by about a factor of $\sim 3$ \citep{xie2019}. This motivates us to
use a model appropriate to MAD systems when attempting to estimate the electron heating parameter $\delta$.

This paper is organized as follows. In Sec.\ \ref{sec:obs}, we provide the observational and modeling results for Sgr A* and M87*, with an emphasis on the accretion rate $\mdotbh$, the bolometric lumonosity $L_{\rm bol}$, and the radiative efficiency $\epsilon$. In Sec.\ \ref{sec:model_efficiency}, we present theoretical calculations of the radiative efficiency of MAD flows corresponding to different values of $\delta$. We then compare these theoretical results with the values of $\epsilon$ inferred by the EHTC for Sgr A* and M87*, and we thereby constrain the value of $\delta$. The final section is devoted to a brief summary.

\section{Observational Constraints on the Radiative Efficiency}\label{sec:obs}

Both M87* and Sgr A* are low-luminosity systems with hot accretion flows. They are the main targets of the EHT project. Below we provide basic properties of the two sources and derive their radiative efficiencies.

\subsection{M87*}\label{sec:m87}

M87* is located at a distance of $d=16.9$ Mpc, and the BH mass is measured to be $\mbh = 6.2\times10^{9}\msun$ \citep{gebhardt2011, eht2019}. By comparing the values of four physical quantities obtained from theoretical predictions and those derived from the reconstructed EHT image of M87* and ALMA-only (Atacama Large Millimeter/submillimeter Array) measurements, the EHTC concluded that the accretion flow in M87* is very likely in a MAD state \citep{eht2021}. This result has been confirmed recently by \citet{yuan2022} by comparing the predicted Faraday Rotation Measure (RM) in M87* with that measured along the jet. The advantage of the latter work is that it does not suffer from uncertainties in the electron temperature and the contribution of nonthermal electrons in the accretion flow, which are a challenge for the EHTC analysis. The spin of the BH in M87* is hard to determine \citep{eht2021}; but some recent works have begun to constrain its value by comparing observed images of the M87* jet with theoretical predictions \citep{co2022, yang2022}. 

Broad-band quasi-simultaneous observations of M87* indicate that the turnover frequency of the sub-mm bump in the $\nu L_\nu$ plot is located at $\sim 230$ GHz (\citealt{algaba2021}, see also \citealt{hada2011}), and that the bolometric luminosity is approximately $L_{\rm bol}\approx8.5\times10^{41}\,\ergs \approx 1.1\times10^{-6}\, \ledd$. In estimating the bolometric luminosity, we only include nuclear emission, cf. Fig.~16 and Model 1a of \citet{algaba2021}\footnote{Note that \citet{prieto2016} derived a larger value of the bolometric luminosity, $L_{\rm bol}\approx 3\times10^{42}\ergs$. The main difference from \citet{algaba2021} is the treatment of data in optical/UV.}. We adopt a $20\%$ uncertainty in $L_{\rm bol}$. 

The second piece of information we need is the mass accretion rate in M87*. From the RM measurements of the nucleus of M87 at sub-millimeter wavelengths, and using an argument previously developed by \citet{agol2000} and \citet{quataert2000}, \citet{kuo2014} derived an upper limit on the accretion rate: $\mdotbh < 9.2\times10^{-4} \msunpyr$. EHT observations now allow a more careful analysis.

The one-zone model discussed in \citet{eht2019} provides a first approximate estimate of $\mdotbh$. The emission size (radius $R$) of the accretion flow is directly measured from the diameter of the ring in the EHT image. Combining this with the sub-mm flux, brightness temperature, and synchrotron peak frequency, one estimates the density as well as the magnetic field strength at radius $R$. This, along with an estimate of the radial velocity of the gas, then gives an approximate estimate of the mass accretion rate, $\mdotbh \sim 25\times10^{-4} \msunpyr$.

The above estimate is further refined in \citet{eht2021}, where EHT polarimetric data are included as important additional constraints, and elaborate theoretical models are developed based on GRMHD simulations. The conclusion of this work is that the mass accretion rate near the BH horizon in M87* lies in the range $\mdotbh\approx (3-20)\times10^{-4}\msunpyr$ \citep{eht2021}, or equivalently, $\mdotbh/\medd = (2.2-15)\times10^{-6}$. This is the range we use in the present work. Note that the EHT result is generally consistent with the initial estimate of \citet{kuo2014} based on polarimetric data.

\begin{figure*}
\centering
\vspace{0.5cm}
\includegraphics[width=0.7\textwidth]{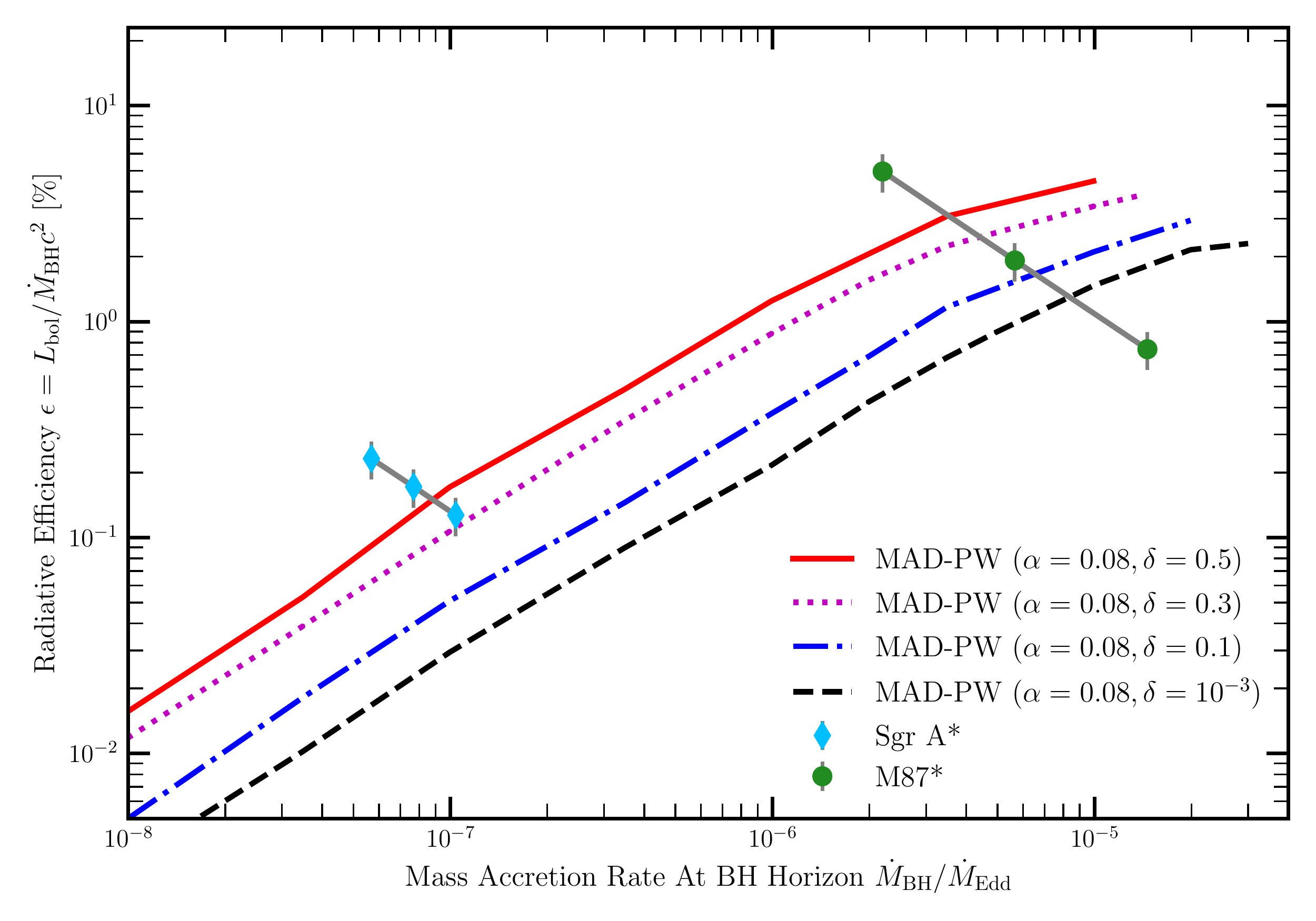}
\caption{Radiative efficiencies $\epsilon$ vs mass accretion rates $\mdotbh/\medd$ of MAD systems. Observational constraints obtained from M87* and Sgr A* (see Sec.\ \ref{sec:obs}) are shown, respectively, by the green circles and cyan diamonds, connected by gray solid curves. The colored lines correspond to numerical results from a theoretical, pseudo-Newtonian (Paczy\'nski-Wiita) potential version of MAD. From top to bottom, the curves correspond to different values of the electron heating fraction: $\delta = 0.5$ (solid red), 0.3 (dotted purple), 0.1 (dot-dashed blue), $10^{-3}$ (dashed black). For a given $\delta$, the radiative efficiency in the models scales as $\epsilon \propto \mdotbh^{0.92}$. All calculations use a viscosity parameter $\alpha = 0.08$ (see Sec.\ \ref{sec:alpha}).}\label{fig:eff}
\end{figure*}

Combining the estimates of the bolometric luminosity and mass accretion rate discussed above, we derive the radiative efficiency $\epsilon$ (cf. Equation\ \ref{eq:eff}) of M87* to be
\begin{equation}
 7.5\times 10^{-3} \la \epsilon \la 5.0\times10^{-2},   \label{eq:m87}
\end{equation} 
where the range is driven almost entirely by the uncertainty in $\mdotbh$. The two extreme values of the radiative efficiency and their geometric mean are shown in Figure\ \ref{fig:eff} as green filled circles. The gray solid curve connecting the three points represents the allowed combinations of $\epsilon$ and $\mdotbh$.

We note that the above estimates of $\mdotbh$ are based on the assumption that the electrons have a thermal distribution. Although electrons and ions in a hot accretion flow are poorly coupled via Coulomb collisions, electron-electron collisions are more effective and will maintain a thermal distribution, at least for moderately large accretion rates. \citet{mahadevan1997} have shown that synchrotron self-absorption couples electrons efficiently at even lower accretion rates. They estimate that electrons remain thermal so long as
\begin{equation}
\frac{\mdotbh}{\medd} > 10^{-4}\,\frac{\alpha^2}{\Theta_e},
\end{equation}
where $\alpha\sim 0.1$ is the effective viscosity parameter of the accretion flow (\citealt{shakura1973}, see also Sec. \ref{sec:alpha}) and $\Theta_e = kT_e/m_ec^2 \sim$\ few is the dimensionless electron temperature in the hot accretion flow. M87* satisfies the above constraint, so the thermal assumption is probably reasonable.

A weak nonthermal tail in the electron distribution is partially equivalent to a thermal model with a higher electron temperature (e.g., \citealt{eht2022a}). However, models that include a significant non-thermal component are quite different from thermal models \citep{eht2019}. For a given radio flux, including non-thermal electrons generally leads to a lower estimate for the accretion rate (lower optical depth), and thus to a larger radiative efficiency.

\subsection{Sgr A*}\label{sec:sgra}

The second object we consider is Sgr A*. It is at a distance of $d=8.13$ kpc and has a BH mass of $\mbh = 4.14\times10^{6}\msun$ (both are average values, as adopted in \citealt{eht2022a}). According to the EHT work, the accretion flow in Sgr A* is probably in a MAD state; specifically, \cite{eht2022a} showed that MAD models are more consistent with observational constraints in Sgr A* compared to SANE models, although the favored models are generally too variable compared to observations (a presently unsolved problem).

The spectrum of Sgr A* peaks at the sub-millimeter waveband \citep{vf2018, bower2019,eht2022b}, and the luminosity of this sub-mm bump is $\approx 5.0\times10^{35}\,\ergs$ (cf. \citealt{bower2019,eht2022b}). The near-infrared emission \citep{witzel2018}, which also originates from the inner accretion flow, should be included when we estimate the bolometric luminosity. Adopting a spectrum that takes the form of a power-law with exponential cutoff near $10^{13}$ Hz (i.e., $F_\nu \propto \nu^\alpha \exp(-\nu/10^{13}\,{\rm Hz})$, \citealt{bower2019}), we empirically derive a luminosity of $\approx 1.9\times10^{35}\,\ergs$ for the infrared emission (we take the 50th percentile of \citet{witzel2018} as representative, see their Figure 19). Combining the sub-millimeter and infrared contributions, the bolometric luminosity of Sgr A* is estimated to be $L_{\rm bol}\approx 6.9\times10^{35}\,\ergs = 1.5\times10^{-9}\,\ledd$. This model-independent measurement of $L_{\rm bol}$ agrees with values derived from MAD models \citep{eht2022a}, which are in the range $L_{\rm bol}=(6.8-9.2)\times10^{35}\,\ergs$. The X-ray emission of Sgr A* during the quiescent state  mainly originates from the region $R>10^4 R_{\rm g}$ (here $R_{\rm g} = G\mbh/c^2$ is the gravitational radius of the BH). Detailed analysis of high-resolution {\it Chandra} observations indicates that nuclear ($R<10^3 R_{\rm g}$) X-ray emission has only $\nu L_{\nu}\approx1.0\times10^{32}\,\ergs$ at 5 keV \citep{roberts2017}. It is three orders of magnitude fainter than the sub-millimeter bump, thus we ignore this contribution. For the uncertainty in our $L_{\rm bol}$ measurement of Sgr A*, we again assume that it is $20\%$ of $L_{\rm bol}$.

Early hot accretion flow models of Sgr A* invoked mass accretion rates at the BH comparable to the Bondi accretion rate, but such large rates were thrown into doubt when linear polarization was reported in Sgr A* at millimeter wavelengths by \citet[][later confirmed by \citealt{bower2003}]{aitken2000}. In two influential papers, \citet{agol2000} argued that this detection implied that RM must be low and hence $\mdotbh$ must be $<10^{-8}\msunpyr$, while \citet{quataert2000} used a similar argument to estimate $\mdotbh \approx 10^{-8}\msunpyr$.

Recently, \citet[][see references therein for earlier theoretical/observational studies]{eht2022a} combined observational constraints from the 230 GHz EHT image of Sgr A* with detailed simulations-based theoretical models, to come up with a tight constraint on the mass accretion rate in Sgr A*: $\mdotbh = (5.2-9.5)\times10^{-9}\,\msunpyr$, or equivalently $\mdotbh / \medd = (5.7-10.4)\times 10^{-8} $. Although these models do not include any spatially resolved polarization constraints from the EHT (none have been published yet), the estimated $\mdotbh$ is perfectly consistent with the original estimates from \citet{agol2000} and \citet{quataert2000}, and subsequent work following the same lines. We use the quoted EHT-derived range in the present work.

Using the above estimates of the bolometric luminosity and the mass accretion rate, we estimate the radiative efficiency $\epsilon$ of Sgr A* to be in the range
\begin{equation}
 1.3\times 10^{-3} \la \epsilon \la 2.3\times10^{-3}.   \label{eq:sgra}
\end{equation}
This range is shown in Figure\ \ref{fig:eff} as cyan diamonds connected by a gray solid curve. The constraint on $\epsilon$ in the case of Sgr A* is much tighter than in M87*. This is because of the factor of $<2$ uncertainty in $\mdotbh$, which is perhaps a little optimistic. However, none of our conclusions will be affected even if we allow another factor of 2 uncertainty.

\subsection{Comparing the Two Objects}

The radiative efficiencies of M87* and Sgr A*, together with the large difference in their Eddington-scaled accretion rates, confirm an important theoretical prediction of hot accretion flow theory. All models, however much they may differ in physics details or parameter choices, agree that the radiative efficiency of hot accretion flows should decrease with decreasing Eddington-scaled mass accretion rate $\mdotbh/\medd$ (e.g., \citealt{narayan1995, narayan1998, xie2012, xie2019}, and Figure\ \ref{fig:eff} here). This is clearly borne out by the EHT-derived estimates for M87* and Sgr A*. However, due to the large uncertainty in $\mdotbh/\medd$ (especially the factor of $\sim 7$ uncertainty in the case of M87*), it is not possible to reliably estimate the slope of the $\epsilon - \mdotbh/\medd$ relation from current observational data. 

\section{Radiative Efficiency of the MAD Model}\label{sec:model_efficiency}

\subsection{The MAD model}\label{sec:model}

In our numerical model of hot accretion flows in the MAD state, we assume for simplicity a non-spinning BH and adopt a pseudo-Newtonian Paczy\'nski-Wiita gravitational potential \citep{PW1980} to mimic the potential of the BH. Since relativity is not explicitly taken into account in the dynamics of our model, the radial velocity can unphysically exceed the speed of light $c$ near the BH horizon. To correct for this, we follow \citet{xie2010} and interpret the velocity $V_{r,{\rm PW}}$ in our Paczy\'nski-Wiita potential model as $\gamma V_{r,{\rm real}}$, where $\gamma$ is the bulk Lorentz factor of the accreting gas and $V_{r,{\rm real}}$ is the corrected radial velocity. 

The neglect of BH spin in the model is reasonable. Although the jet power depends sensitively on spin (e.g., \citealt*{tchekhovskoy2010}), the emission from the main body of the accretion flow has only a weak dependence. This is because, unlike the cold accretion disk model \citep{shakura1973} where the location of the disk inner edge varies substantially with BH spin, a hot accretion flow extends down to the BH horizon $R_{\rm horizon}$ which is much less sensitive to BH spin. Dynamical properties of the flow outside $(3-5)R_{\rm horizon}$ are mostly unaffected by BH spin, while emission from regions inside of $\sim 2R_{\rm horizon}$ (e.g., radius of marginally bound orbit) is mostly beamed toward the BH and lost through the horizon.

The details of our MAD model and its radiation calculation can be found in \citet{xie2019}. Below we highlight the main points. Guided by numerical simulations, \citet{xie2019} solve in cylindrical coordinates the height-integrated equations of a steady MAD system, paying attention to the effects of non-axisymmetric gas streams/spirals, which are a dominant feature of hot accretion flows in the MAD regime (e.g., \citealt{narayan2003, tchekhovskoy2011, mckinney2012, white2019, chatterjee2022}). We follow the GRMHD simulations of \citet{mckinney2012} to set a $R^{-s_{bz}}$ profile of the global vertical component of the magnetic field $B_z$, where $s_{bz}\approx 1.1$. The field strength is determined by the azimuthally-averaged plasma $\beta$ (gas-to-magnetic pressure ratio) parameter at the outer boundary $200 R_{\rm g}$, i.e. $\bar{\beta}_{z0}=1.5$ (see \citealt{xie2019} for the definition of $\bar{\beta}_{z0}$). Other components are calculated numerically according to the gas dynamics, where the magnetic Prandtl number $\mathcal{P}_{\rm m}=2$ and a parameter $\kappa_\phi = -0.5$ are introduced respectively, for the calculation of $B_r/B_z$ and $B_\phi/B_z$. With this setup (strength and profile) of global magnetic fields, the vertical magnetic flux threading the accretion flow inside $10\,R_{\rm g}$ is about $50\,(\mdotbh R_{\rm g}^2 c)^{1/2}$, in agreement with that threading the BH in numerical simulations of MAD systems (\citealt{tchekhovskoy2011, mckinney2012, davis2020, narayan2022}, see their reported values of the parameter $\phi_{\rm BH}$). The stress of global ordered magnetic fields, approximately proportional to $B_z B_\phi$  \citep{xie2019}, can then be derived. Besides the global field, we also include a turbulent magnetic field, by setting the gas-to-turbulent-magnetic pressure ratio $\beta_{\rm t}=10$. Most importantly, unlike most MHD simulations where a single-fluid equation is adopted, we use separate energy equations for the electrons and ions \citep{narayan1995, yuan2014,xie2019}. As usual, in our model we mimic the turbulent stress term, which is automatically present in MHD simulations, via a radius-independent viscosity parameter $\alpha$ (see Sec.\ \ref{sec:alpha} below for the determination of its value).

Numerical simulations suggest that hot accretion flows, including MADs, have a strong mass outflow (e.g., \citealt{narayan2012, yuan2012, yuan2015, yang2021}) outside of a certain radius $R_{\rm flat}$ ($\sim 6-10 R_{\rm g}$; refer to Fig. 5 in \citealt{yuan2015} and Fig. 6 in \citealt{yang2021}). Inside $R_{\rm flat}$, the outflow is highly suppressed and the accretion rate is nearly a constant. Instead of the conventional broken power-law fit, we adopt a smooth function for $\dot{M}(R)$, viz., $\dot{M}(R) = \mdotbh \left[1+ ({R-R_{\rm BH})^2/R^2_{\rm flat}}\right]^{s/2}$, where $s$ measures the strength of the mass outflow. This expression ensures that $\dot{M}(R)\propto R^{s}$ for $R\gg R_{\rm flat}$ and $\dot{M}(R)\approx \mdotbh$ for $R\ll R_{\rm flat}$. In this work we follow the BH spin $a=0$ MAD case of \citet{yang2021} and set $R_{\rm flat}\approx 6R_{\rm g}$, $s\approx0.2$. Since the outflow strength we adopt is fairly weak (e.g., $\dot{M}(200 R_{\rm g})\approx 2\mdotbh$), it has only a minor impact on the radiative efficiency of the model.

\subsection{Viscous parameter $\alpha$ for the turbulent stress term} \label{sec:alpha}

The ``viscous parameter'' $\alpha$ \citep{shakura1973}, which describes the strength of the turbulent stress, is a free parameter in our model. We note that MHD turbulence can be automatically generated and sustained in MHD simulations of accretion flows (e.g., \citealt{stone2001,igumenshchev2003,white2019}), thus $\alpha$ is implicitly included there. In our model, $\alpha$ has a strong effect on the predicted radiative output of the model. The radiation emitted by a hot accretion flow is largely determined by the gas density $\rho$ which, for a given mass accretion rate, is determined  by the efficiency of angular momentum transport. The latter is proportional to $\alpha$ in the SANE case (note that in our MAD model the stress by global ordered magnetic fields is handled separately, based on numerical simulation results, cf. Sec.\ \ref{sec:model}). For a SANE model, we typically have $\rho \propto \mdotbh V_{r}^{-1}\propto \alpha^{-1}\mdotbh$ \citep{narayan1994}. A similar expression can also be applied for the magnetic field strength.\footnote{We note that for the low-$\dot{M}/\medd$ regime appropriate to a hot accretion flow, most radiation is in the form of synchrotron, and for optical depth $\tau\la 10^{-6}$ inverse Compton scattering only plays a negligible role. In this case, the radiative luminosity scales as $L_{\rm bol}\propto \rho B^{2} \propto\rho^2 \propto\alpha^{-2}\mdotbh^{2}$ (assuming that the electron temperature is un-affected if $\epsilon \ll 1$, cf. \citealt{narayan1998} and Figure\ \ref{fig:eff}).} In addition, the total dissipated energy is determined by the strength of turbulence, i.e. it is linearly proportional to $\alpha$ in our model, cf. \citet{xie2019}. Since part of the dissipated energy goes into electrons (through the $\delta$ parameter), the value of $\alpha$ has a direct and strong impact on the radiative luminosity.

\begin{figure*}
\centering
\vspace{0.5cm}
\includegraphics[width=0.7\textwidth]{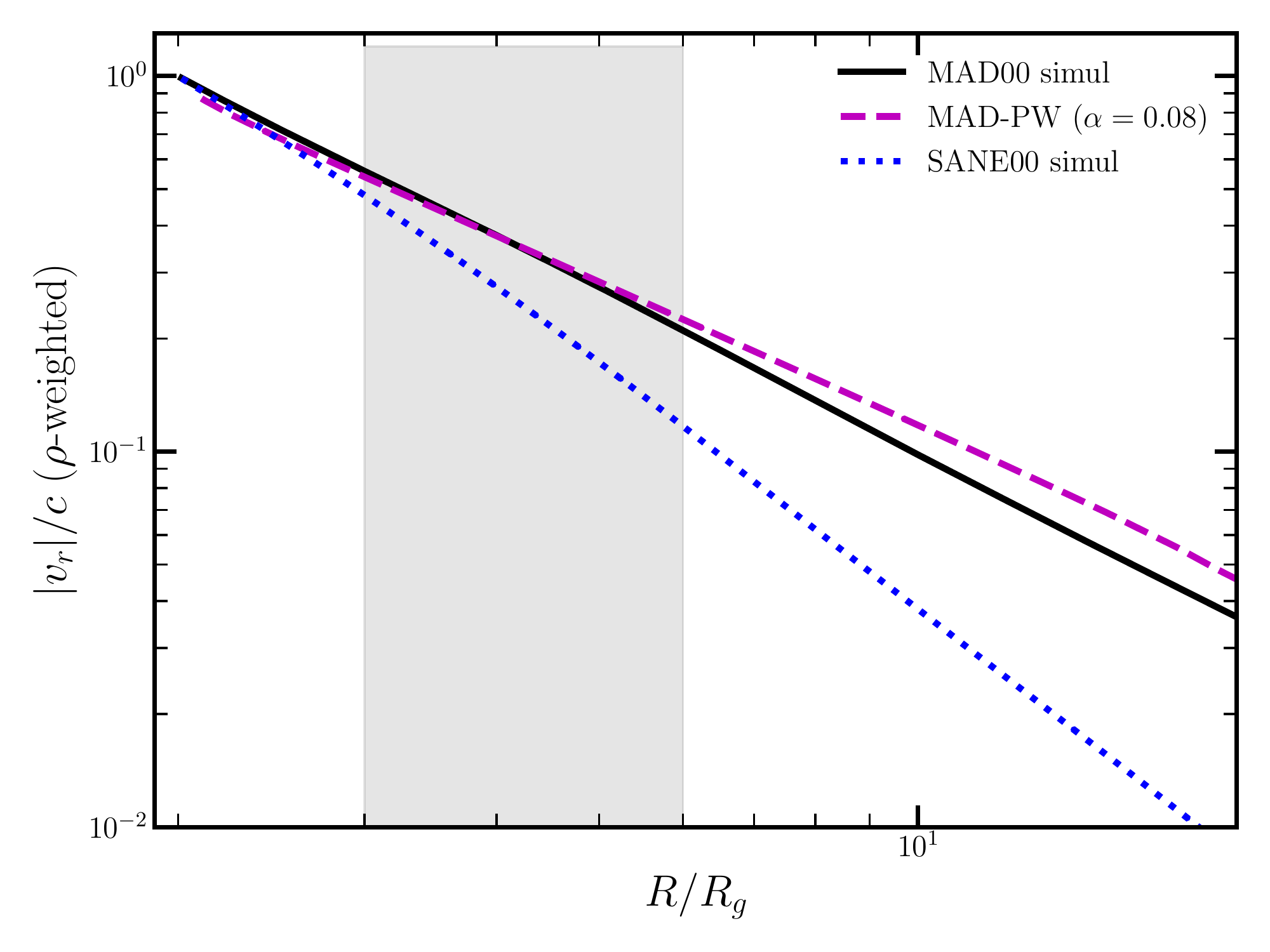}
\caption{Estimation of the viscosity parameter $\alpha$ corresponding to the turbulent stress. The black solid and blue dotted curves show the density-weighted radial velocity in GRMHD simulations of MAD and SANE respectively, around a non-spinning BH \citep{narayan2022, ricarte2022}. The purple dashed curve is derived from our height-integrated MAD model with $\alpha=0.08$. The shadowed region ($3R_{\rm g} < R < 6R_{\rm g}$) shows where most of the observed radiation comes from.}\label{fig:alpha_est}
\end{figure*}

As shown in Figure\ \ref{fig:alpha_est}, we use the radial velocity measured in a GRMHD MAD simulation \citep{narayan2022} of a non-spinning BH to calibrate the value of $\alpha$ in our model. We focus on the region $R \approx (3-6)R_g$ from which most of the observed synchrotron radiation in the EHT band (230\,GHz) comes, shown by the shadowed region in Figure\ \ref{fig:alpha_est}.

The calculated radial profile of the radial velocity as obtained from our height-integrated MAD model with $\alpha = 0.08$ is shown as the purple dashed line in Figure\ \ref{fig:alpha_est}. In comparison, the black solid curve shows the velocity profile seen in a long-duration GRMHD MAD simulation with BH spin $a=0$ \citep{narayan2022}. The two profiles agree very well, especially in the shaded region which produces most of the radiation observed by the EHT. Therefore, we use $\alpha=0.08$ in computing all our model results.  For completeness, we also show by the blue dotted curve the GRMHD simulation result for a SANE model with $a=0$ \citep{ricarte2022}. There is a large difference, suggesting that the global magnetic field in MAD plays an important role in transferring angular momentum and speeding up the radial velocity.

\subsection{Radiative efficiency} \label{sec:deltae}

We now calculate the radiative efficiency of our height-integrated MAD model using various values of the electron heating fraction parameter $\delta$. As mentioned earlier, there are two source of heating for electrons in a hot accretion flow: one is through energy transfer from ions via electron-ion Coulomb collisions,
and the other is through direct viscous heating at a rate proportional to $\delta$ (see equation~\ref{eq:energy}). The  Coulomb heating rate is $\propto \rho^{2}$, whereas the viscous heating rate is $\propto \rho$ \citep[e.g.,][]{narayan1995}. Since $\rho \propto \mdotbh$, we expect the impact of $\delta$ to be most evident at low values of $\mdotbh/\medd$. Thus, among the two sources we are considering, Sgr A* is the most likely to give a useful constraint on $\delta$. 

For our calculations, we do not focus on detailed microphysics of viscous dissipation (see {\it Introduction}). Instead, we take a ``model-independent'' approach in which we assume a constant electron heating fraction $\delta$ that is independent of radius and explore the observational constraint on its value. We adopt several values: $\delta =10^{-3}, ~0.1, ~0.3$, and $0.5$. The calculated radiative efficiencies corresponding to these values of $\delta$ are shown as a function of $\mdotbh/\medd$ in Figure\ \ref{fig:eff}. In our calculations, we limit ourselves to the ``advection-dominated regime'' which corresponds to low accretion rates \citep{yuan2014}. Both M87* and Sgr A* belong to this regime. The maximal accretion rate at the horizon, above which the radiation is so strong that the flow is no longer advection-dominated but enters into the ``luminous hot accretion flow regime'' \citep{yuan2001,yuan2014}, is determined numerically by having the advection term $f_{\rm adv} = 1-q^-/q_{\rm vis}$ \citep{narayan1994} to be zero at any radius. In our model, this limit is located at $\sim (1-3)\times10^{-5}\medd$. 

Because the strong poloidal field in a MAD system breaks axisymmetry and causes the accretion to occur in the form of dense gas streams surrounded by dilute magnetic voids \citep{narayan2003, tchekhovskoy2011, mckinney2012, white2019, chatterjee2022}, at a given accretion rate, the density of the gas in the accreting streams in MAD will be higher than that of the more uniform density in SANE. Consequently, the ion-electron Coulomb heating becomes comparable to the turbulent heating at a lower value of $\mdotbh/\medd$ in a MAD model compared to a SANE model (compare the results shown in Fig.\ \ref{fig:eff} with those in \citealt{xie2012}).

As seen in Figure\ \ref{fig:eff}, the maximal accretion rate of the ADAF regime in MAD models (the upper limits of the curves) decreases with increasing $\delta$. This is naturally expected. At a given accretion rate, a higher $\delta$ means a larger fraction of the viscously dissipated energy goes to electrons. This makes the electrons hotter, and radiatively more luminous (higher in radiative cooling $q^-$). Consequently, it will reach a lower advection term $f_{\rm adv}$.

In addition, for a typical ADAF regime with $\mdotbh/\medd \la 6\times10^{-6}$, we find that the radiative efficiency has a roughly linear dependence on the accretion rate: $\epsilon \propto \mdotbh^{0.92}$ (for a similar result, see also \citealt{xie2019}). Equivalently, the bolometric luminosity follows $L_{\rm bol}\propto \mdotbh^{1.92}$. We note that an estimation of $L_{\rm bol} \propto \mdotbh^{2}$ was derived previously in \citet*{narayan1998} for the SANE case.

We now explore the $\delta$-dependence of $\epsilon$ based on our calculations. We find that, at a given accretion rate, the radiative efficiency differs by a factor of $\sim 5$ for different values of $\delta$. Besides, when $\delta\la 0.05$, the radiative efficiency becomes insensitive to the value of $\delta$, because electron heating by Coulomb collisions with ions is dominant. For the typical ADAF accretion rate regime with $\mdotbh/\medd \la 6\times 10^{-6}$, we have $\epsilon \propto {\rm max}(\delta^{0.7},0.1)$, i.e. $\delta=0.5$ and $\delta=0.3$ cases are brighter by a factor of $\approx 3$ and $\approx2.2$ respectively, than the $\delta=0.1$ case. Combining the dependence on $\delta$ and $\mdotbh$, our model results can be summarized as,
\begin{eqnarray}
\epsilon & \approx & 1.9\%\left({\dot{M}_{\rm BH}\over10^{-6}\medd}\right)^{0.92}\,{\rm max}(\delta^{0.7},0.1), \nonumber\\
& & \quad {\rm where}\,\,\, \left({\dot{M}_{\rm BH}\over10^{-6}\medd}\right) < 6.
\end{eqnarray}
The dependence of $\epsilon$ on $\delta$ becomes weaker when $\mdotbh/\medd \ga 6\times 10^{-6}$, with $\epsilon \approx (2-4)\%$. This is because of the increasing importance of Coulomb collision heating of electrons at high accretion rates. 

\subsection{Constraint on $\delta$ from Sgr A* and M87*}

Finally, we apply the above theoretical calculations to the observational constraints from Sgr A* and M87*. From Figure\ \ref{fig:eff} we find that in the case of Sgr A*, low values of $\delta$ (i.e. $\delta<0.3$) are ruled out, and that $\delta\sim 0.5$ is preferred, given the uncertainties in $\mdotbh$. 
On the other hand, we cannot put any constraint on $\delta$ based on the current observational data from M87*. A useful constraint at a similar level to Sgr A* will require the mass accretion rate in M87* to be further constrained to $\mdotbh/\medd \la 8\times10^{-6}$ (equivalently, $\mdotbh \la 10^{-3} \msun\,{\rm yr}^{-1}$), i.e., the upper limit on $\mdotbh$ will need to be reduced by a factor $\sim2$ compared to what the EHT has achieved so far.

\section{Summary}
In the theory of hot accretion flows around black holes, an important but poorly understood parameter is $\delta$, which describes the fraction of viscous energy that directly heats electrons (refer to Equation \ref{eq:energy}). The value of $\delta$ determines the thermal energy of electrons and therefore the radiative efficiency of the accretion flow for a given accretion rate. While the underlying microphysics is complicated and the value of $\delta$ is poorly determined from theory, in this paper we try to constrain its value by using the most recent EHTC observational and modeling results on M87* and Sgr A*. 

EHTC papers have provided good constraints on the mass accretion rates at the BH horizon for the two sources. These results, combined with the measured bolometric luminosities of the two objects, lead to useful constraints on the radiative efficiencies, as presented in Equations (\ref{eq:m87}) and (\ref{eq:sgra}) and shown in Figure\ \ref{fig:eff}.
Meanwhile, we can analytically solve the height-integrated dynamical equations of a hot accretion flow in the MAD regime and calculate the theoretically expected radiative efficiency. For this calculation, we need to assume a value for the viscous parameter $\alpha$. By comparing the radial velocity profile in the height-integrated model with the velocity profile obtained in numerical GRMHD simulations, we estimate that $\alpha\sim0.08$. Using this $\alpha$, we calculate the radiative efficiency in our model as a function of the accretion rate and the electron heating fraction parameter $\delta$. By comparing these theoretical results with the observational constraints on Sgr A* and M87* from EHTC, we are able to estimate the value of $\delta$.

In the case of Sgr A*, we find that we can rule out $\delta \la 0.3$, and that the most likely value is $\delta \sim 0.5$. This result is in excellent agreement with that obtained by \citet{yuan2003} who modeled the spectral energy distribution of Sgr A*. The analysis in the present paper was made possible because the EHTC reported a tight constraint on the mass accretion rate in Sgr A*. Even if we doubled their uncertainty range, our results would still be largely unchanged. However, we caution that our height-integrated model of MAD accretion makes several approximations whose effects are difficult to quantify.

In the case of M87*, we are unable to obtain a useful constraint on $\delta$ because EHTC does not provide  a sufficiently strict constraint on the mass accretion rate.

\begin{acknowledgements}
FGX is supported in part by National SKA Program of China (No. 2020SKA0110102), the National Natural Science Foundation of China (NSFC; grants 11873074, 12192220, and 12192223), and the Youth Innovation Promotion Association of CAS (Y202064). FY is supported in part by NSFC (grants 12133008, 12192220, and 12192223). RN is supported in part by the National Science Foundation under grants OISE-1743747 and AST1816420. RN's work was carried out at the Black Hole Initiative, Harvard University, which is funded by grants from the John Templeton Foundation and the Gordon and Betty Moore Foundation.
\end{acknowledgements}
{}
    
\end{document}